\documentclass[11pt,twoside]{article}

\usepackage{asp2014}

\aspSuppressVolSlug
\resetcounters

\begin{document}

\title{Software Architecture and System Design of Rubin Observatory}



\author{William~O'Mullane,$^1$ Frossie~Economou,$^1$ Kian-Tat~Lim,$^2$ Fritz~Mueller,$^2$ Tim~Jenness,$^1$ Gregory~P.~Dubois-Felsmann,$^3$ Leanne~P.~Guy,$^1$ Ian~S.~Sullivan,$^4$ Yusra~AlSayyad,$^5$ John~D.~Swinbank,$^6$ $^5$ and K.~Simon~Krughoff$^1$}
\affil{$^1$Rubin Observatory Project Office, 950 N.\ Cherry Ave., Tucson, AZ  85719}
\affil{$^2$SLAC National Accelerator Laboratory,  2575 Sand Hill Rd., Menlo Park, CA 94025}
\affil{$^3$IPAC, California Institute of Technology, MS 100-22, Pasadena, CA 91125}
\affil{$^4$University of Washington, Dept.\ of Astronomy, Box 351580, Seattle, WA 98195}
\affil{$^5$Department of Astrophysical Sciences, Princeton University, Princeton, NJ 08544}
\affil{$^6$ASTRON, Oude Hoogeveensedijk 4, 7991\,PD, Dwingeloo, The Netherlands}
\paperauthor{William~O'Mullane}{womullan@lsst.org}{0000-0003-4141-6195}{Rubin Observatory Project Office}{}{Tucson}{AZ}{85719}{USA}
\paperauthor{Frossie~Economou}{}{0000-0002-8333-7615}{Rubin Observatory Project Office}{}{Tucson}{AZ}{85719}{USA}
\paperauthor{Kian-Tat~Lim}{}{0000-0002-6338-6516}{SLAC National Accelerator Laboratory}{}{Menlo Park}{CA}{94025}{USA}
\paperauthor{Fritz~Mueller}{}{0000-0002-7061-4644}{SLAC National Accelerator Laboratory}{}{Menlo Park}{CA}{94025}{USA}
\paperauthor{Tim~Jenness}{}{0000-0001-5982-167X}{Rubin Observatory Project Office}{}{Tucson}{AZ}{85719}{USA}
\paperauthor{Gregory~P.~Dubois-Felsmann}{}{0000-0003-1598-6979}{IPAC}{}{Pasadena}{CA}{91125}{USA}
\paperauthor{Leanne~P.~Guy}{}{0000-0003-0800-8755}{Rubin Observatory Project Office}{}{Tucson}{AZ}{85719}{USA}
\paperauthor{Ian~S.~Sullivan}{}{0000-0001-8708-251X}{University of Washington}{}{Seattle}{WA}{98195}{USA}
\paperauthor{Yusra~AlSayyad}{}{}{Department of Astrophysical Sciences}{}{Princeton}{NJ}{08544}{USA}
\paperauthor{John~D.~Swinbank}{}{0000-0001-9445-1846}{Department of Astrophysical Sciences}{}{Princeton}{NJ}{08544}{USA}
\paperauthor{K.~Simon~Krughoff}{}{0000-0002-4410-7868}{Rubin Observatory Project Office}{}{Tucson}{AZ}{85719}{USA}

\begin{abstract}
Starting from a description of the Rubin Observatory Data Management System Architecture, and drawing on our experience with and involvement in a range of other projects including Gaia, SDSS, UKIRT, and JCMT, we derive a series of generic design patterns and lessons learned.
\end{abstract}

\section{Introduction}

The Legacy Survey of Space and Time (LSST) \citep{2019ApJ...873..111I} is a ``wide fast deep'' optical/near-IR survey of half the sky in \emph{ugrizy} bands to a combined depth of \emph{r} $\sim$ 27.5 (36\,nJy) based on 825 visits over ten years.
Carried out by the Vera C.\ Rubin Observatory (Rubin) on Cerro Pach\'{o}n (with an altitude of 2647\,m) in Chile, the survey will produce around 100\,PB of data consisting of about a billion 16\,Mpix images, enabling measurements for 40 billion objects.
Rubin Observatory will take an image approximately every 40\,s (slew and settle time plus 30\,s exposure time), which leads to around 20\,TB of images streaming off the mountain from Chile each night.
Rubin's LSST is not the first wide-field imaging survey, but the combination of depth, area, and throughput makes it uniquely challenging.

The observatory is due to go into full operations in late 2024.
In the meantime, we routinely operate the Rubin Auxiliary Telescope with the LSST Atmospheric Transmission Imager and Slitless Spectrograph (LATISS) instrument as both an imager and spectrograph \citep{2020SPIE11452E..0UI}.
During regular operations, it will be used as a spectrograph to measure atmospheric transmission, but during construction, it has been used as an imager to integrate and commission the data management system.
For regularly updated key milestones, see \citet{DMTN-232}.

In this paper, we introduce the vision, architecture, and guiding values of the Rubin Data Management (DM) System. We then discuss some lessons learned building the DM system, including comparisons to other projects, primarily Gaia.

\section{System Vision}

The mission statement for Rubin Data Management (DM) is to ``Stand up operable, maintainable, quality services to deliver high-quality LSST data products for science and education, all on time and within reasonable cost.'' DM will deploy its software for producing and serving the data products, and the elements of the DM system include the following.

DM will transfer the images from Chile to the US data facility, SLAC, within seven seconds by using the 100\,Gbps long-haul network, which was an early investment of the project.
Once at SLAC, the images are processed in parallel through the prompt processing system (see Section \ref{sec:prompt}), which, within minutes, distributes alerts of astronomical sources which have moved, changed, or appeared.
After 80 hours, the images will be available to the data rights holders.
On a roughly annual cadence, DM will reprocess all the images taken since the start of operations and release new catalogs and other products as defined in \citet{LSE-163} (see Section \ref{sec:DRP}).

\subsection{Democratizing research in astronomy}

We will also serve these data products on the Rubin Science Platform.
Because data at this volume does not fit on a laptop, we provide the infrastructure for researchers to bring their code to the data rather than the data to their code.
Because the load is on our servers, users need only an internet connection and a browser to allow for sophisticated experimentation.

In this way, the Rubin Science Platform provides a level playing field for interacting with Rubin data.
Open-source software and open data are essential to open science and reproducibility. However, open data alone is not sufficient for inclusivity.
We must also find ways to support researchers who are resource-poor (lacking the computing resources associated with major research universities), time-poor (have a high teaching load, few/no grad students or postdocs), or who work in liberal arts colleges, historically black colleges, or other places that lack an extensive peer network for technical and research support.
Lowering the barrier to entry requires minimizing the investment (time, money, experience) necessary to meaningfully engage with the scientific questions that can be resolved with the data.

\section{Architecture}

Figure \ref{fig:arch} shows a simplified view of the system architecture. The full details are publicly available in \citep{LDM-148}. All the DM code is available on GitHub at \url{https://github.com/lsst}.

\begin{centering}
\articlefigure[width={9.5cm}]{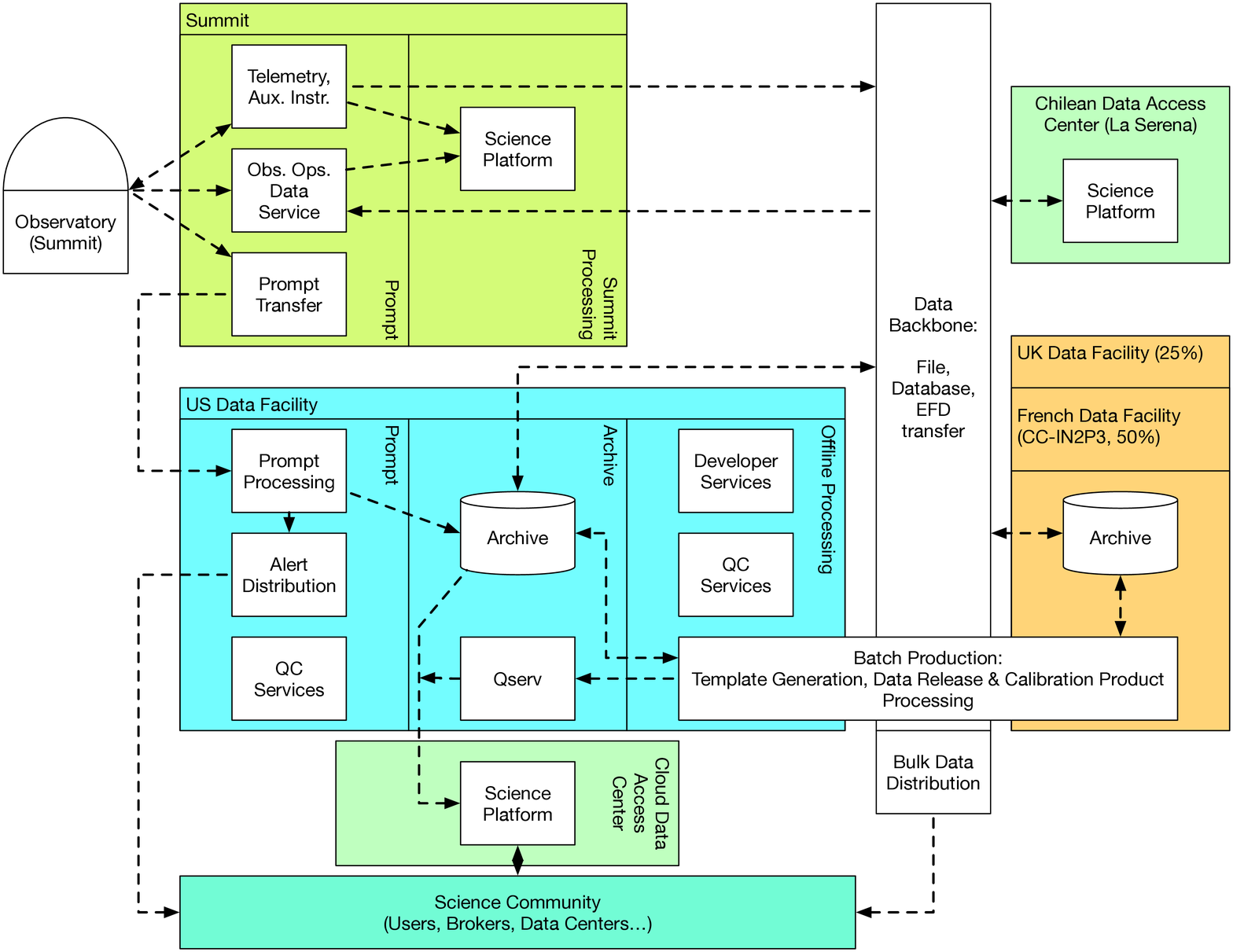}{fig:arch}{ Simplified Vera C. Rubin Architecture diagram from Magic Draw }
\end{centering}

DM's work commences once the image is read out of the Camera. DM already gathers some information that the Camera software puts in the image header to make a minimally meaningful image.
As the data is written, the Camera software also provides a quick look at the image.
Once available, the image is written to the Observatory Operations Data Service (OODS) and simultaneously transferred to the US Data Facility (USDF) at SLAC via the Prompt Transfer System. Though we use Rucio for transfers between facilities, the Prompt Transfer System requires custom code to support faster transfers.

On the summit, a restricted access Science Platform allows staff to interact with the images in the OODS directly.
A cluster of about 400 cores is available for quick ad-hoc processing in situ, but we expect most processing to be done at the USDF.

\subsection{Prompt Processing} \label{sec:prompt}
The Prompt Processing framework runs at the USDF. However, many of the framework's components will be reused to drive rapid analysis and quick-look functionality at the Summit and test stand facilities.
The design of Prompt Processing is driven by the requirement that alerts be distributed within 120 seconds of completion of the readout of the last exposure of the visit.
To enable as much I/O and computation as possible to be done in advance, we instantiate one process per CCD when the summit sends a next\_visit Kafka event.
These next\_visit events provide notice of the telescope pointing, exposure duration, filter selection, and other metadata at least 20 seconds in advance of the first exposure of a visit.
Upon receiving the next\_visit event, we use \texttt{knative} in a Kubernetes environment to prepare a new container where we connect to the Butler to pre-load reference catalogs, calibration products, templates, solar system ephemerides, and prior alert history.
Once the raw images corresponding to an earlier next\_visit event for a given detector finish downloading to a local Ceph object store, the images are ingested to the container-local Butler, and the Alert Production pipeline payload begins processing.
The Alert Production pipeline produces packaged alerts streamed to the Community Brokers, writes all data products to the repository at the USDF with the Butler, and updates the Alert Pipeline Database (APDB) with new measurements.
Detailed information on the initial design and prototype in the Google Cloud environment can be found in \citet{DMTN-219}.
Figure \ref{fig:pp} provides a flow chart for prompt processing.

\begin{centering}
\articlefigure[width={5.5cm},angle=-90]{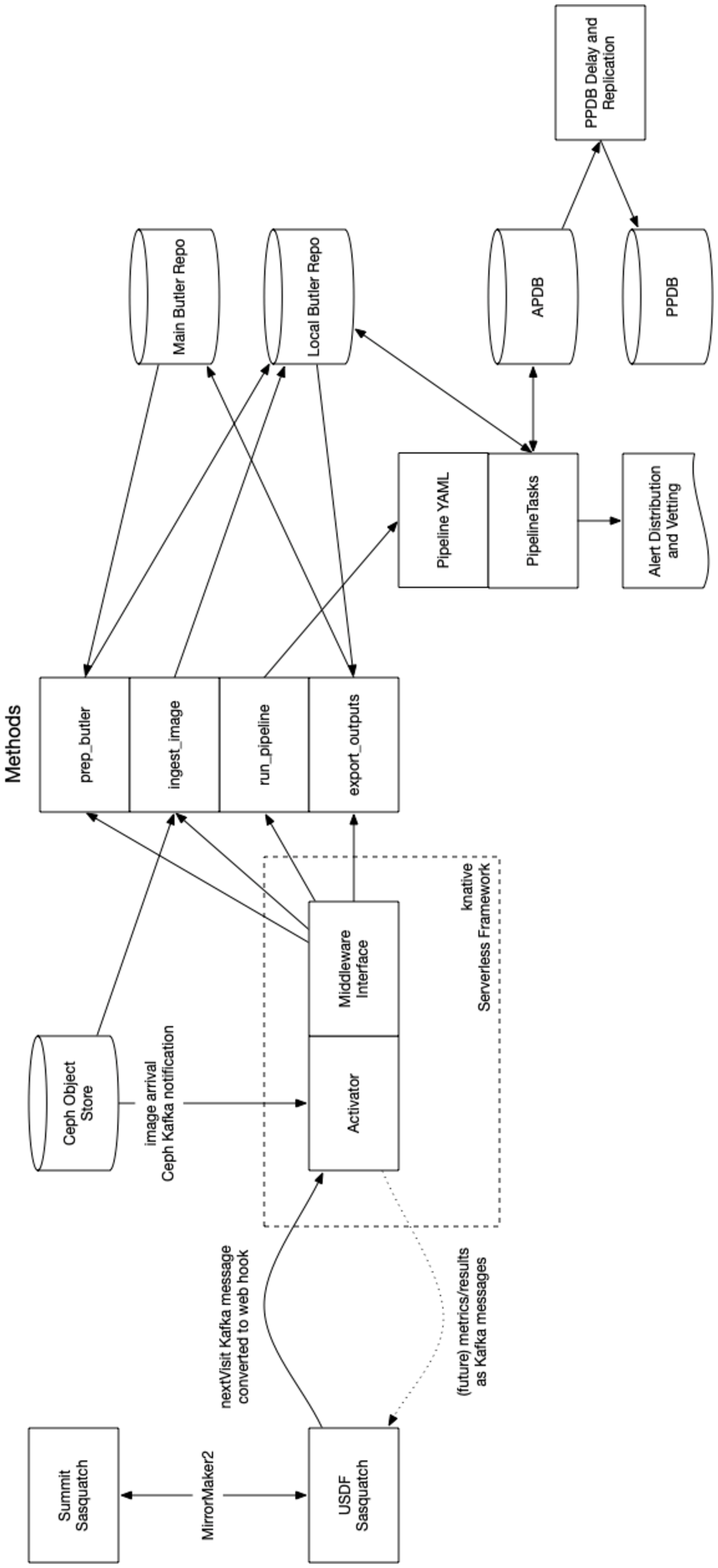}{fig:pp}{ Prompt Processing flow diagram.}
\end{centering}

\subsection{Data Release Processing}\label{sec:DRP}
About once per year, we will reprocess all data from the start of the survey.
We will spend a few months performing pilot runs and validation before starting the nine-month batch processing.
Individual jobs are distributed between France, the US, and the UK data facilities using PanDA \citep{DMTN-213}
The batch processing system is described by \citet{P52_adassxxxii} in more detail in this issue.
The US data facility distributes the quantum graphs and raw images for execution at the three sites (figure \ref{fig:drp}).
Copies of the resulting catalogs and processed images will be available from all three data facilities.

\begin{centering}
\articlefigure[width={6cm},angle=-90]{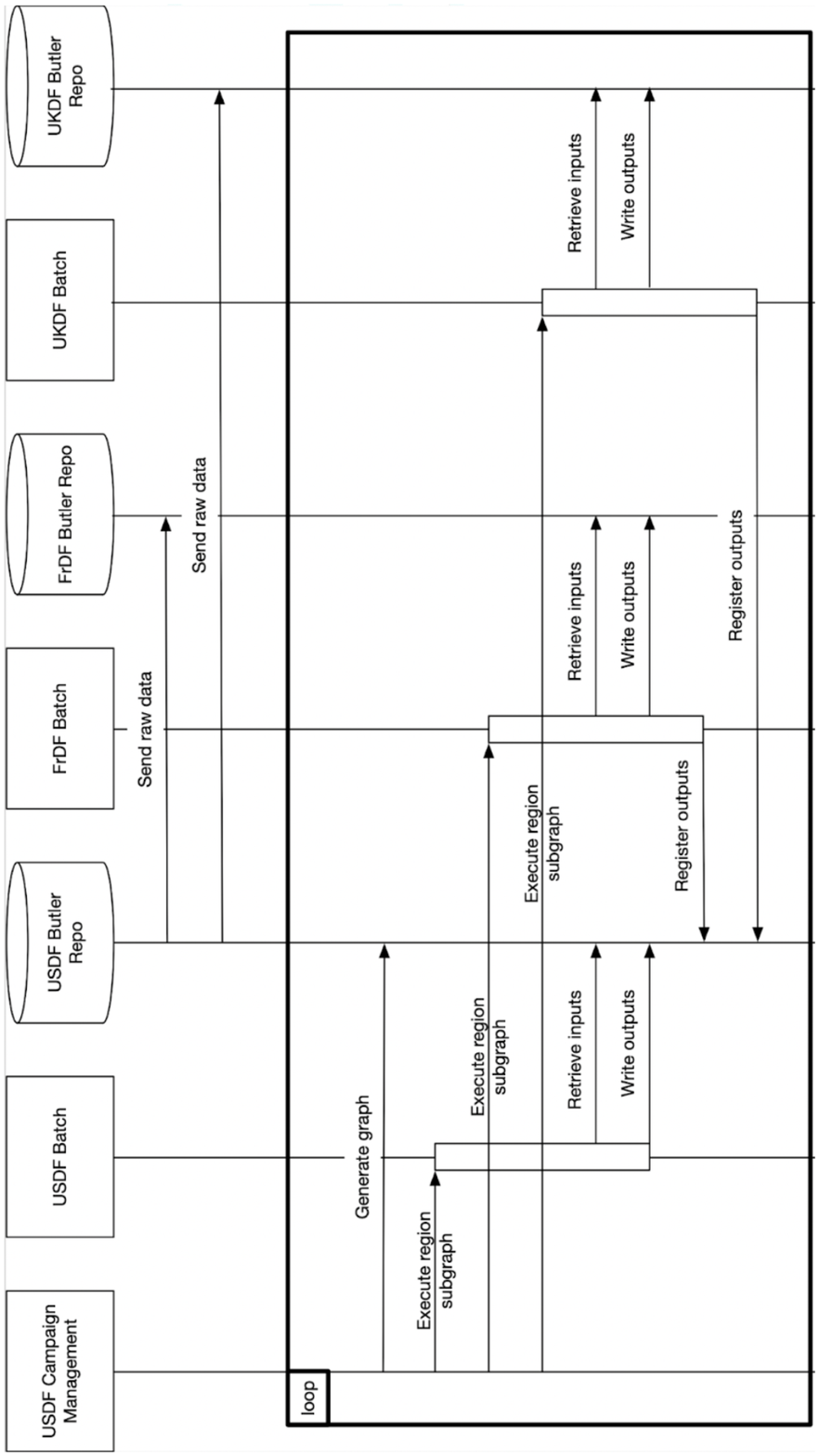}{fig:drp}{Data Release Production event chart showing communication between the US, French, and UK data facilities (USDF, FrDF, and UKDF respectively). Butler repos are described by \citet{C24_adassxxxii}.}
\end{centering}

\subsection{Data Access} \label{sec:dataaccess}

Data rights holders will have access to the catalogs and images via the Rubin Science platform.
Catalogs will be accessible via relational databases and column stores to support different access patterns.
The object table contains the measurements that have utilized all the visits and is estimated at $4 \times 10^{10}$ rows.
However, the ForcedSource table, which contains the lightcurves (one row for each observation of each object), will run to $10^{12}$ rows.
Qserv \citep{C15_adassxxxii} was built specifically to answer astronomy queries quickly for large numbers of users.
Because astronomers often ask for many rows, table scans are often unavoidable.
Qserv's shared-nothing distributed architecture can serve multiple queries with single shared table scans.

We have compared Qserv's performance to non-relational databases such as Google's BigQuery \citet[e.g.,][]{Document-31100} and column stores such as parquet.
Qserv provides the best value for catalog queries.
However, we see the advantage of non-relational technology for
some \emph{unpredictable} and \emph{complex} access that Qserv may or may not handle for user-defined functions, pattern matching, or unusual iteration schemes.
For example, we expect users to compute a statistic on every lightcurve using a small number of columns.
Many cloud tools, such as Spark and DASK, work best with column stores like parquet.
Therefore, we plan to make parquet files available to serve this access pattern and as a backup for Qserv.


Because compute is cheap, but storage is expensive on the cloud, we have chosen a hybrid model.
We will hold most data on premises at SLAC but run the science platform in the cloud \citep{2021arXiv211115030O}.
The ingress is free, and the user egress is manageable.

\subsection{Cloud Native}
\label{sec:cloudnative}
Like many projects \citep{2017ASPC..512...33O}, Rubin leaned heavily on containers early on.
We also quickly understood the need for sophisticated container orchestration and settled on Kubernetes (K8S).
This decision drove service architectures that are well isolated from the underlying infrastructure.
This approach has already paid massive dividends:

\begin{itemize}
\item When funding lines suddenly shifted, we were able to painlessly transition from an on-premises facility to an Interim Data Facility on Google Cloud.
\item The Rubin Science Platform (RSP) became a generic data services platform that is currently deployed on eight distinct (and distinctly managed) infrastructures (on-prem and cloud).
\item Cloud can now be freely leveraged for services, like the RSP, which benefit from its advantages, such as elasticity, scalability, and isolation.
\end{itemize}

Our architectural approach is geared towards lowering the cost of developing and deploying a new data service.
Services utilize a common infrastructure (Phalanx\footnote{\url{http://phalanx.lsst.io/}}) providing services such as authentication and authorization, secrets management, Transport Layer Security (TLS) certificates, and templates to speed up the creation of new services in the FastAPI framework.
The GitOps infrastructure for K8S deployment using ArgoCD takes care of easy per-infrastructure configuration and deployment.

On the summit in Cerro Pach\'{o}n, where we have on-prem machines, the Chile DevOps team uses Foreman and Puppet to bring up a full K8S infrastructure.
Both DM and Telescope and Site software then deploy services on top of this infrastructure.
Deploying control components on K8S allows for better resilience.

\section{Lessons learned}
We would like to share some observations from this project in several areas.
\subsection{Standards}
Standards are excellent. They minimize the learning curve for new hires, which is a major problem on all software projects.
Gaia used the European Cooperation for Space Standardization \citep[ECSS;][]{2008ASPC..394..191O}, and
 Rubin used Model-Based Systems Engineering \citep[MBSE;][]{2018SPIE10705E..0US}.

We also have the International Virtual Observatory (IVOA) standards for astronomy. Gaia archive is fully IVOA based \citep{2019ASPC..523..445S,2015scop.confE...8G}, and the Sloan Digital Sky Survey (SDSS) implemented and helped define many of the original protocols \citep{2005ASPC..347..684T}.
Rubin is IVOA first, with implementations of TAP, HiPS, and the SODA cutout service.
Rubin uses DataLink to abstract image access from ObsTAP results, which is further utilized within the system to expose IVOA services and use them internally.

One upside of picking IVOA standards is that many implementations are now available.
Rubin uses the CADC's TAP implementation with our own Qserv plugin; users of the TAP service have no idea they are using Qserv.
That allows us to use Firefly for visualization fairly easily, as it is
fully VO based.

\subsection{Architecture}
There is a lot of analysis and design to develop an architecture -- standards help with that.
Undoubtedly tools to support your chosen standards are handy in the beginning. Later they may become cumbersome. Both Rubin and Gaia started with Rational Architect and switched to Magic Draw.
Rubin still maintains a complete set of requirements and design elements in Magic Draw, while Gaia switched to code as prime and reverse engineers some diagrams for documentation purposes.

For verification, Rubin uses Jira Test Manager \citep{2018SPIE10705E..0US} while Gaia used an in-house system based on open software \citep{2012SPIE.8449E..0GC}.
A systematic and automated approach to verification is needed from the outset.

One difficulty on both projects was writing clear written and testable requirements. In hindsight, both projects could have done better.
We can easily fall into the trap of assuming that all agree on vague statements or requirements when usually a little delving will show quite the opposite.
It is worth putting in the effort early to write how we think things will work in detail and as precisely as possible.
This precision requires systems engineering, which is often underestimated.

Regardless of the project, systems tend to be split up at the outset into  $7 \pm 2$ subsystems.
Frequently large projects start all subsystems together, but often starting each subsystem as needed would work better.
For example, on Gaia, Coordination Unit 1 (CU1 Architecture) and CU3 (Astrometry) were concentrated initially, with other Units trailing by some months.
People from CU1 and CU3 could move to those other units.
CU9 (Gaia Archive) was purposefully delayed until launch was close.
On Rubin, all DM work breakdown structure elements started together.
Not all components were needed initially.
However, the simultaneous start led to good engagement, reflected in the developer guide and project management approach.

\subsection{Documentation}
We found that a good document publishing and indexing system is essential.
We recommend providing templates for standard documents early on to make it easy for people to follow the standards. texmf works well for \LaTeX. Both Rubin and Gaia have "bibfile" generation for all recorded docs.

Most Gaia docs used the provided Latex templates and were in SVN, while Livelink held the published PDFs, and the Livelink search system worked to some extent.
Metadata in Livelink was curated, and only documentalists were allowed to upload documents.
The Livelink API allowed for the easy construction of bibfiles.

Rubin developed a documentation infrastructure that further lowers the barrier to documentation by providing templated creation via bespoke Slackbot. It uses the same IDE/toolchain developers use for coding, supports Restructured Text and \LaTeX, and publishes via GitHub.
Single page documents, technotes, and site-based documentation (e.g., \url{https://pipelines.lsst.io}) share the same infrastructure \citep{SQR-000} and search indexing hub (\url{https://www.lsst.io}).
This system has made documents easy to find, remember and edit in one's preferred IDE.
In contrast, the Rubin project beyond DM stores change-controlled documents (PDFs and Word) in Docushare, which has a search function that is difficult to use.
Its frequent failure to find documents may be because it relies on metadata which is often incomplete or incorrect.  Furthermore, we failed to crack the Docushare API for the automatic generation of bibfiles.

We also recommend building project glossaries early, maintaining them, and promoting their use.
Both Rubin and Gaia also have tools to generate acronym lists from documents (text or tex -- not Word).\footnote{\url{http://gaia.esac.esa.int/gpdb/glossary.txt}, \url{https://www.lsst.org/scientists/glossary-acronyms}}. Gaia goes one step further with all constants used in docs and code stored in a single parameter database \citep{2005ESASP.576...67D}, which requires work from the outset.

\subsection{Interfaces}
Rubin DM defines and maintains many interfaces.
The first interface \emph{separates the data model from the persistence mechanism}, which implements one of our core tenets.
The Rubin Butler ensures algorithms never access data directly \citep{2022SPIE12189E..11J,2019ASPC..523..653J,C24_adassxxxii,P52_adassxxxii}.
The Butler passes Python Objects to clients with algorithm code unaware of data location or file formats.
Similarly, Gaia had data trains and the Main DataBase (MDB) dictionary, which insulated algorithms from data access since the outset \citep{1999BaltA...8...57O}.

The interfaces between systems and others are controlled by Interface Requirements Documents (IRD) and Interface Control Documents (ICD).
These need system engineering and test plans from the outset.
What Rubin calls ICDs are only IRDs.
The Gaia MDB dictionary holds all data models \citep{2015ASPC..495...47H, 2011ASPC..442..351O} and is the basis of the ICD between the subsystems.
This dictionary was insufficient, and we had to work later to make data transfers and processing work well.

\subsection{Products, repositories and technology stacks}

Not all projects use product trees, but they can be very useful.
The Rubin product tree identified all software products and who was responsible for them in construction.
This product tree was good but infrequently updated (partially, perhaps, because it was in MagicDraw).
The product tree should help group packages into products and clarify dependencies.
However, on Rubin, we have hundreds of repos on GitHub, the dependencies are not straightforward, and few repos are usable as standalone packages.
This means that we need to build a complete set from source, yielding a
 mono-build.
 While we have a package-based set of GitHub repos, which is the correct pattern, getting cleanly defined buildable packages has proved difficult.
The use of Conda environments has allowed improvements, but it started late.
We discontinued patched versions for third-party packages by ensuring that corresponding packages existed in conda-forge.
The middleware has been made independent and put on PyPI, allowing other projects, such as SPHEREx, to adopt it.
Some software best practices are given in \citet{2018SPIE10707E..09J}.

Gaia has a huge SVN repo of everything based on the product tree. Builds are done on parts of the SVN tree -- dependencies strictly managed at Jar level via Nexus.
Printed-out full product trees are impressive to see the amount of work to do and are useful at early reviews.

\subsection{Deployment}
As mentioned in \S\ref{sec:cloudnative} we are cloud-native on Rubin.
Abstracting infrastructure effectively (Kubernetes / container orchestration, middleware) facilitates wider adoption of software and services by others, reducing context-switching penalties and supporting continuing expertise.
At a low level, DM uses Puppet, but SLAC was already using Chef and continue to do so --using both of these is unfortunate on one project.

Gaia chose Java for portability and ease of coding.
There were no containers, but Jars were always deployed from Nexus.
All configurations for various machines were in SVN deployment scripts that pulled the correct versions to a specific machine.

\subsection{Databases}
Some people love them, and some people hate them, some of us love them and hate them, but databases are a part of any big system, and choosing the correct one is hard.
We think databases are great for persistence, the ability to query in different ways, and relational support for Atomicity, Consistency, Isolation, and Durability (ACID).
Nevertheless, there are problems with centralization/replication, schema evolution, and performance cliffs.
We have difficulties with multi-user/multi-tenant systems but
REST APIs in front helps a lot.

A common mistake is trying to use one single database.
More is better, and per-application databases, sometimes specialized (Redis, InfluxDB), add resilience and are more manageable nowadays.
On Rubin, we have InfluxDB for summit Engineering, Postgres for observing logs and ancillary info, and AlertsDB.

We use Cassandra for Prompt Products, and of course, we have developed Qserv in-house for catalog access \citep{C15_adassxxxii}.
Gaia had at least Intersystems Cache for processing \citep{2011ExA....31..215O}, Postgres for archive, and the dictionary.

\subsection{Open software project management}
It seems appropriate to mention management here though that could be a topic for a full paper or book of its own
(many insights may be found in \citet{OMULLANE2005}).
First, leadership is required for complex astronomy/software projects, not just management.
Finding good leaders is very hard, requiring domain expertise and management training.
We help by spreading some management across several people, exposing them to the issues in the large project and ways to deal with them.
Most importantly, provide support to potential managers/leaders.
One must acknowledge that this route is not for everyone -- experimenting is good but gives people a route back in a short timescale if they decide it is not what they wanted.

While it is possible to do Agile, the NSF and other agencies require earned value \citep{2014SPIE.9150E..1EG,2016SPIE.9911E..0NK} to help understand what is being delivered.
We recommend finding good managers who understand technical and managerial needs, which is difficult.
Of course, the aim is to build a techno/scientific culture in leadership and breed more managers of the required ilk and
create community and collaboration around a codebase nurtured by those managers.
To help
we should be offering opportunities for getting career credit for supporting the mission and its community, not just first-to-publish.

Running a big project also requires many agreements with institutions.
To build open-source software, we recommend putting it in the contracts/agreements from the outset but without being overly explicit about the license.
Licensing is important.
Rubin picked GPL at the outset but now prefers a less restrictive license.
We have found it very hard to change at this stage. .

\acknowledgments This material or work is supported in part by the National Science Foundation through Cooperative Agreement AST-1258333 and Cooperative Support Agreement AST1836783 managed by the Association of Universities for Research in Astronomy (AURA), and the Department of Energy under Contract No. DE-AC02-76SF00515 with the SLAC National Accelerator Laboratory managed by Stanford University.

\bibliography{I08}

\begin{thebibliography}{}
\expandafter\ifx\csname natexlab\endcsname\relax\def\natexlab#1{#1}\fi
\expandafter\ifx\csname url\endcsname\relax
  \def\url#1{\texttt{#1}}\fi
\expandafter\ifx\csname urlprefix\endcsname\relax\def\urlprefix{URL }\fi
\providecommand{\eprint}[2][]{\url{#2}}

\bibitem[{{Comoretto} et~al.(2012){Comoretto}, {Gallegos}, {Els}, {Gracia},
  {Lock}, {Mercier}, \& {O'Mullane}}]{2012SPIE.8449E..0GC}
{Comoretto}, G., {Gallegos}, J., {Els}, S., {Gracia}, G., {Lock}, T.,
  {Mercier}, E., \& {O'Mullane}, W. 2012, in Modeling, Systems Engineering, and
  Project Management for Astronomy V, vol. 8449 of Proc.\ SPIE, 84490G

\bibitem[{{de Bruijne} et~al.(2005){de Bruijne}, {Lammers}, \&
  {Perryman}}]{2005ESASP.576...67D}
{de Bruijne}, J.~H.~J., {Lammers}, U., \& {Perryman}, M.~A.~C. 2005, in The
  Three-Dimensional Universe with Gaia, edited by {C.~Turon, K.~S.~O'Flaherty,
  \& M.~A.~C.~Perryman}, vol. 576 of ESA Special Publication, 67

\bibitem[{{Gill} et~al.(2014){Gill}, {Gracia}, {Lupton}, \&
  {O'Mullane}}]{2014SPIE.9150E..1EG}
{Gill}, R., {Gracia}, G., {Lupton}, R.~H., \& {O'Mullane}, W. 2014, in
  Modeling, Systems Engineering, and Project Management for Astronomy VI, vol.
  9150 of Proc.\ SPIE, 91501E

\bibitem[{{Gonzalez-Nunez}(2015)}]{2015scop.confE...8G}
{Gonzalez-Nunez}, J. 2015, in Science Operations 2015: Science Data Management,
  8

\bibitem[{Gower et~al.(2023)}]{P52_adassxxxii}
Gower, M., et~al. 2023, in ADASS XXXII, edited by S.~{Gaudet}, S.~{Gwyn},
  P.~{Dowler}, D.~{Bohlender}, \& A.~{Hincks} (San Francisco: ASP), vol. TBD of
  ASP Conf.\ Ser., TBD

\bibitem[{{Hernandez} \& {Hutton}(2015)}]{2015ASPC..495...47H}
{Hernandez}, J., \& {Hutton}, A. 2015, in ADASS XXIV, edited by A.~R. {Taylor},
  \& E.~{Rosolowsky}, vol. 495 of ASP Conf.\ Ser., 47

\bibitem[{{Ingraham} et~al.(2020)}]{2020SPIE11452E..0UI}
{Ingraham}, P., et~al. 2020, in Software and Cyberinfrastructure for Astronomy
  VI, vol. 11452 of Proc.\ SPIE, 114520U

\bibitem[{{Ivezi{\'c}} et~al.(2019)}]{2019ApJ...873..111I}
{Ivezi{\'c}}, {\v Z}., et~al. 2019, \apj, 873, 111. \eprint{arXiv:0805.2366}

\bibitem[{{Jenness} et~al.(2022){Jenness}, {Bosch}, {Salnikov}, {Lust},
  {Pease}, {Gower}, {Kowalik}, {Dubois-Felsmann}, {Mueller}, \&
  {Schellart}}]{2022SPIE12189E..11J}
{Jenness}, T., {Bosch}, J.~F., {Salnikov}, A., {Lust}, N.~B., {Pease}, N.~M.,
  {Gower}, M., {Kowalik}, M., {Dubois-Felsmann}, G.~P., {Mueller}, F., \&
  {Schellart}, P. 2022, in Software and Cyberinfrastructure for Astronomy VII,
  vol. 12189 of Proc.\ SPIE, 1218911. \eprint{arXiv:2206.14941}

\bibitem[{Jenness et~al.(2018)}]{2018SPIE10707E..09J}
Jenness, T., et~al. 2018, in Software and Cyberinfrastructure for Astronomy V,
  vol. 10707 of Proc.\ SPIE, 1070709

\bibitem[{{Jenness} et~al.(2019)}]{2019ASPC..523..653J}
{Jenness}, T., et~al. 2019, in ADASS XXVII, edited by P.~J. {Teuben}, M.~W.
  {Pound}, B.~A. {Thomas}, \& E.~M. {Warner}, vol. 523 of ASP Conf.\ Ser., 653.
  \eprint{arXiv:1812.08085}

\bibitem[{Juri\'{c} et~al.(2021)}]{LSE-163}
Juri\'{c}, M., et~al. 2021, {Data Products Definition Document}.
  \urlprefix\url{https://lse-163.lsst.io/}

\bibitem[{{Kantor} et~al.(2016){Kantor}, {Long}, {Becla}, {Economou}, {Gelman},
  {Juric}, {Lambert}, {Krughoff}, {Swinbank}, \& {Wu}}]{2016SPIE.9911E..0NK}
{Kantor}, J., {Long}, K., {Becla}, J., {Economou}, F., {Gelman}, M., {Juric},
  M., {Lambert}, R., {Krughoff}, S., {Swinbank}, J.~D., \& {Wu}, X. 2016, in
  Modeling, Systems Engineering, and Project Management for Astronomy VI,
  edited by G.~Z. {Angeli}, \& P.~{Dierickx}, vol. 9911 of Proc.\ SPIE, 99110N

\bibitem[{Lim(2022{\natexlab{a}})}]{DMTN-213}
Lim, K.-T. 2022{\natexlab{a}}, {DMTN-213: Multi-Site Data Release Processing
  Using PanDA and Rucio}. {Vera C. Rubin Observatory},
  \urlprefix\url{https://dmtn-213.lsst.io/}

\bibitem[{Lim(2022{\natexlab{b}})}]{DMTN-219}
--- 2022{\natexlab{b}}, {DMTN-219: Proposal and Prototype for Prompt
  Processing}. {Vera C. Rubin Observatory},
  \urlprefix\url{https://dmtn-219.lsst.io/}

\bibitem[{Lim et~al.(2018)Lim, Bosch, Dubois-Felsmann, Jenness, Kantor,
  O'Mullane, \& Petravick}]{LDM-148}
Lim, K.-T., Bosch, J., Dubois-Felsmann, G., Jenness, T., Kantor, J., O'Mullane,
  W., \& Petravick, D. 2018, {LDM-148: Data Management System Design}.
  \urlprefix\url{https://LDM-148.lsst.io/}

\bibitem[{Lust et~al.(2023)}]{C24_adassxxxii}
Lust, N., et~al. 2023, in ADASS XXXII, edited by S.~{Gaudet}, S.~{Gwyn},
  P.~{Dowler}, D.~{Bohlender}, \& A.~{Hincks} (San Francisco: ASP), vol. TBD of
  ASP Conf.\ Ser., TBD

\bibitem[{Mueller et~al.(2023)}]{C15_adassxxxii}
Mueller, F., et~al. 2023, in ADASS XXXII, edited by S.~{Gaudet}, S.~{Gwyn},
  P.~{Dowler}, D.~{Bohlender}, \& A.~{Hincks} (San Francisco: ASP), vol. TBD of
  ASP Conf.\ Ser., TBD

\bibitem[{{O'Mullane}(2005)}]{OMULLANE2005}
{O'Mullane}, W. 2005, {Large Scientific Data Systems: analysis of some existing
  projects and their applicability to Gaia}, Tech. rep., University of
  Barcelona. Treball GAIA-C1-ESAC-HA-WOM-003,
  \urlprefix\url{https://dms.cosmos.esa.int/COSMOS/doc_fetch.php?id=497678}

\bibitem[{O'Mullane(2022)}]{DMTN-232}
O'Mullane, W. 2022, {Celebratory Milestones}.
  \urlprefix\url{https://dmtn-232.lsst.io/}

\bibitem[{{O'Mullane} et~al.(2021){O'Mullane}, {Economou}, {Huang}, {Speck},
  {Chiang}, {Graham}, {Allbery}, {Banek}, {Sick}, {Thornton}, {Masciarelli},
  {Lim}, {Mueller}, {Padolski}, {Jenness}, {Krughoff}, {Gower}, {Guy}, \&
  {Dubois-Felsmann}}]{2021arXiv211115030O}
{O'Mullane}, W., {Economou}, F., {Huang}, F., {Speck}, D., {Chiang}, H.-F.,
  {Graham}, M.~L., {Allbery}, R., {Banek}, C., {Sick}, J., {Thornton}, A.~J.,
  {Masciarelli}, J., {Lim}, K.-T., {Mueller}, F., {Padolski}, S., {Jenness},
  T., {Krughoff}, K.~S., {Gower}, M., {Guy}, L.~P., \& {Dubois-Felsmann}, G.~P.
  2021, in ADASS XXI, ASP Conf.\ Ser., in press. \eprint{arXiv:2111.15030}

\bibitem[{{O'Mullane} et~al.(2008){O'Mullane}, {Hoar}, \&
  {Lammers}}]{2008ASPC..394..191O}
{O'Mullane}, W., {Hoar}, J., \& {Lammers}, U. 2008, in ADASS XVII, edited by
  R.~W. {Argyle}, P.~S. {Bunclark}, \& J.~R. {Lewis}, vol. 394 of ASP Conf.\
  Ser., 191. \eprint{arXiv:0712.0249}

\bibitem[{{O'Mullane} et~al.(2011{\natexlab{a}}){O'Mullane}, {Lammers}, \&
  {Hernandez}}]{2011ASPC..442..351O}
{O'Mullane}, W., {Lammers}, U., \& {Hernandez}, J. 2011{\natexlab{a}}, in ADASS
  XX, edited by {I.~N.~Evans, A.~Accomazzi, D.~J.~Mink, \& A.~H.~Rots}, vol.
  442 of ASP Conf.\ Ser., 351

\bibitem[{{O'Mullane} et~al.(2011{\natexlab{b}}){O'Mullane}, {Lammers},
  {Lindegren}, {Hernandez}, \& {Hobbs}}]{2011ExA....31..215O}
{O'Mullane}, W., {Lammers}, U., {Lindegren}, L., {Hernandez}, J., \& {Hobbs},
  D. 2011{\natexlab{b}}, Experimental Astronomy, 31, 215.
  \eprint{arXiv:1108.2206}

\bibitem[{{O'Mullane} \& {Lindegren}(1999)}]{1999BaltA...8...57O}
{O'Mullane}, W., \& {Lindegren}, L. 1999, Baltic Astronomy, 8, 57

\bibitem[{{O'Mullane} et~al.(2017){O'Mullane}, {Morris}, \&
  {Hoar}}]{2017ASPC..512...33O}
{O'Mullane}, W., {Morris}, D., \& {Hoar}, J. 2017, in ADASS XXV, edited by
  N.~P.~F. {Lorente}, K.~{Shortridge}, \& R.~{Wayth}, vol. 512 of ASP Conf.\
  Ser., 33

\bibitem[{{Salgado} et~al.(2019){Salgado}, {Gonz{\'a}lez-Nu{\~n}ez}
  et~al.}]{2019ASPC..523..445S}
{Salgado}, J., {Gonz{\'a}lez-Nu{\~n}ez}, et~al. 2019, in ADASS XXVII, edited by
  P.~J. {Teuben}, M.~W. {Pound}, B.~A. {Thomas}, \& E.~M. {Warner}, vol. 523 of
  ASP Conf.\ Ser., 445

\bibitem[{{Selvy} et~al.(2018){Selvy}, {Roberts}, {Reuter}
  et~al.}]{2018SPIE10705E..0US}
{Selvy}, B.~M., {Roberts}, A., {Reuter}, M., et~al. 2018, in Modeling, Systems
  Engineering, and Project Management for Astronomy VIII, edited by G.~Z.
  {Angeli}, \& P.~{Dierickx}, vol. 10705 of Proc.\ SPIE, 107050U

\bibitem[{Sick(2015)}]{SQR-000}
Sick, J. 2015, {SQR-000: The LSST DM Technical Note Publishing Platform}. {Vera
  C. Rubin Observatory SQuaRE Technical Note},
  \urlprefix\url{https://sqr-000.lsst.io/}

\bibitem[{{Thakar} et~al.(2005){Thakar}, {Szalay}, {O'Mullane}, {Budav{\'a}ri},
  {Nieto-Santisteban}, {Fekete}, {Li}, {Carliles}, {Gray}, \&
  {Lupton}}]{2005ASPC..347..684T}
{Thakar}, A.~R., {Szalay}, A.~S., {O'Mullane}, W., {Budav{\'a}ri}, T.,
  {Nieto-Santisteban}, M.~A., {Fekete}, G., {Li}, N., {Carliles}, S., {Gray},
  J., \& {Lupton}, R. 2005, in ADASS XIV, edited by P.~{Shopbell},
  M.~{Britton}, \& R.~{Ebert}, vol. 347 of ASP Conf.\ Ser., 684

\bibitem[{Thomson(2019)}]{Document-31100}
Thomson, J.~R. 2019, {LSST Benchmarking of Qserv and BigQuery}. {Vera C. Rubin
  Observatory}, \urlprefix\url{http://ls.st/Document-31100}

\end{thebibliography}

\end{document}